\documentstyle[sprocl]{article}
\input{psfig}
\def\Journal#1#2#3#4{{#1} {\bf #2}, #3 (#4)}

\def\PRL{\em Phys. Rev. Lett.}

\def\be{\begin{equation}}
\def\ee{\end{equation}}
\def\bea{\begin{eqnarray}}
\def\eea{\end{eqnarray}}

\def\GeV{{\rm GeV}}
\def \ETmiss{\thinspace\thinspace{\not{\negthinspace\negthinspace E}}_T}
\def \suwgt{{\cal X}}
\def \crit{{\cal C}}
\begin{document}
\title{HOW TO SEARCH FOR A LIGHT STOP AT THE TEVATRON}
\author{ GREGORY MAHLON }
\address{Department of Physics, University of Michigan,
500 E. University Ave., Ann Arbor, MI  48109, USA}
\maketitle\abstracts{
We describe a method for searching for a light stop squark
($\widetilde{M}_t + \widetilde{M}_{LSP} < M_t$) at the Fermilab 
Tevatron.  Traditional searches rely upon stringent
background-reducing cuts which, unfortunately, leave very few signal
events given the present data set.  To avoid this difficulty, we
suggest using a milder set of cuts, combined with a
``superweight,'' whose purpose is to discriminate between signal and
background.  The superweight consists of a sum of terms (each
of which are either zero or one) assigned
event-by-event depending upon the values of various observables.
By construction, the superweight ``large'' for the 
signal and ``small'' for the background.  We apply this method to 
the detection of stops coming from top decay.  It is straightforward
to adapt our method to other processes.}


Motivated by recent suggestive experimental results,\cite{Suggestive}
we consider the detection potential of a light stop
squark at the Fermilab Tevatron, using the current data set.
We focus on SUSY models where the decay 
$ t \rightarrow \tilde{t} \tilde{\chi}_1^0 $
is kinematically allowed.  Furthermore, we assume that the
lightest chargino state, $\tilde\chi_1^{+}$ is heavy enough
to forbid the decay $ \tilde{t} \rightarrow \tilde\chi_1^{+} b$.
In this case, the stop decays via
$ \tilde{t} \rightarrow \tilde\chi_1^0 c$
with 100\% branching fraction.  Thus, the free parameters 
entering into our discussion are $\widetilde{M}_{t}$,
$\widetilde{M}_{\chi_1}$, and 
${\cal{B}}(t\rightarrow\tilde{t}\tilde\chi_1^0)$.

For the purposes of this discussion, we ignore the additional
complications which arise within the framework of a
complete SUSY model ({\it e.g.}\ top decays to $\tilde\chi_2^0$
or $\tilde\chi_3^0$, and additional top production in
the cascade decays of squarks and gluinos \cite{XtraTops}).
Instead, we focus upon the basic signal for the presence
of stops in top quark decay
\begin{eqnarray}
p\bar{p} \rightarrow t\bar{t}  ; &&
t \rightarrow \tilde{t}\tilde\chi_1^0%
\rightarrow c\tilde\chi_1^0 \tilde\chi_1^0 \nonumber \\
&&
\bar{t} \rightarrow \bar{b} W^{-}%
\negthinspace\rightarrow \bar{b}\ell^{-}\bar\nu_{\ell},
\label{Signal}
\end{eqnarray}
(plus the charge-conjugated state), 
which appears in the detector as a charged lepton, two jets,
and missing transverse energy.  The net effect of including the 
complications mentioned above is the appearance of additional soft 
jets in the final state, and, (with the right cuts) a somewhat larger 
signal.\cite{FullWriteup}

An important feature of this process is that
for the stop and LSP masses under investigation, the $\bar{b}$
becomes the jet with the largest transverse energy over 70\%
of the time.  This is a consequence of the larger phase space
for the decay of top to $Wb$ versus its decay to 
$\tilde{t}\tilde\chi_1^0$.  Thus, the properties of the leading jet
are very similar to the properties of the $\bar{b}$ jet.

Assuming tree-level SM production 
with a $K$-factor of unity and a branching ratio of 50\% for
top to stop, we estimate that the cross section times all
branching ratios for (\ref{Signal}) is about 0.6 pb.
This is to be compared to the largest SM background,
the production of a $W$ plus two jets,
which for some set of loose cuts contributes about 770 pb. 
Our goal in introducing the superweight method is to effectively
deal with this large background without cutting away all of the signal.

We begin by requiring each event to pass the series of cuts
designed to reduce the SM backgrounds to a manageable level
(see Table 1).  We then define the superweight $\suwgt$ on
an event-by-event basis as the number of criteria in Table 2 
which are true, that is, 
$
\suwgt \equiv \sum_{i=1}^{N} {\cal C}_i,
$
where each of the ${\cal C}_i$'s evaluate to 0 or 1.
The ${\cal C}_i$ have been chosen such that events coming from the 
signal (Eq. 1) tend to have a large value of $\suwgt$, while 
background events tend to have a small value of $\suwgt$.

There are two issues in the selection of the criteria in Table 2: 
choice of the observable, and choice of the cut point.  
The physics of the signal and backgrounds should  be used as a guide 
in deciding which quantities should be investigated as potential 
superweight elements.   For example, ${\cal C}_9$ was inspired 
by the observation that since the leading jet is usually the $\bar{b}$
jet, it should combine with the observed lepton to form a $\bar{t}$ 
quark.  Hence, there should be an upper limit on the transverse mass 
of the leading jet and the lepton.  We determine the placement of the 
cut point as follows.  First, observe that the mean contribution of 
a given ${\cal C}_i$ to the superweight for some class of events is 
precisely the fraction of events for which that criterion is true.
Put differently, $\langle{\cal C}_i\rangle$ is the area under a plot of 
$(1/\sigma) d\sigma/d{\cal Q}$ lying above ${\cal Q}_0$ for
an observable ${\cal Q}$ and cut point ${\cal Q}_0$.  Thus, to 
select the cut point, we compare the values of this area for the 
signal and background distributions as a function 
of ${\cal Q}_0$, choosing the point where the difference is the 
greatest.  For a superweight criterion to%
\vfill

\noindent
\begin{minipage}[t]{2.30in}
{\footnotesize  Table 1: Cuts applied before
evaluating the superweight. }
\vspace{0.3cm}
\begin{center}
\begin{tabular}{|r@{\thinspace}c@{\thinspace}l|}
\hline
$p_T(\ell)$ & $>$ &  $ 20 \enspace\GeV$ \\
$\ETmiss$ & $>$ &  $ 20 \enspace\GeV$ \\
$p_T(j_h)$ & $>$ &  $ 15 \enspace\GeV \quad\negthinspace\hbox{(jets 1 \& 2)}$\\
$p_T(j_s)$ & $<$ &  $ 10 \enspace\GeV \quad\negthinspace\hbox{(jet 3)}$\\
$\vert\eta(\ell)\vert$ & $<$  & $ 1$ \\
$\vert\eta(j_h)\vert$ & $<$  & $ 2$  \\
$\Delta R(j,j)$  & $>$ & $ 0.4$ \\
$\Delta R(j,\ell)$ & $>$  & $ 0.4$ \\
$m_T(\ell,\ETmiss)$& $>$   & $ 100 \enspace\GeV$ \\
\hline
\end{tabular}
\end{center}
\end{minipage}
\ \
\begin{minipage}[t]{2.30in}
{\footnotesize Table 2:  Superweight components}
\vspace{0.3cm}
\begin{center}
\begin{tabular}{|cr@{\thinspace}c@{\thinspace}l|}
\hline
$\crit_1$ & $\ETmiss$ & $>$ &  $ 65 \enspace\GeV$ \\
$\crit_2$ & $p_T(j_2){+}\ETmiss$ & $>$ &  $ 95 \enspace\GeV$ \\
$\crit_3$ & $\ETmiss{-}p_T(\ell)$ & $>$ &  $ 0 \enspace\GeV$\\
$\crit_4$ & $m_T(\ell,\ETmiss)$ & $>$ & $ 125 \enspace\GeV$\\
$\crit_5$ & $\phi_{j_1,\ell}$ & $<$  & $ 2.4 \enspace{\rm radians}$ \\
$\crit_6$ & $p_T(\ell){+}\ETmiss$ & $>$ &  $ 95 \enspace\GeV$ \\
$\crit_7$ & $p_T(j_1){+}\ETmiss$ & $>$ &  $ 95 \enspace\GeV$ \\
$\crit_8$ & $\cos\theta_{j_1,\ell}$ & $>$  & $ -0.15$ \\
$\crit_9$ & $m_T(j_1,\ell)$ & $<$ & $ 125 \enspace\GeV$\\
$\crit_{10}$ & $m(j_1,j_2,\ell)$& $<$   & $ 200 \enspace\GeV$ \\
\hline
\end{tabular}
\end{center}
\end{minipage}
\eject
\noindent
be useful, the optimal value of ${\cal Q}_0$ should be
reasonably stable over the range of parameters being investigated.
Using the definition in Table 2, we find that the mean superweight
for the signal is 7.4 or greater, depending on the SUSY masses.
For the background, the mean superweight is 2.6.

We present our estimate of the number of signal events 
in 100 pb$^{-1}$ passing all of the cuts which have a superweight 
of 6 or greater in Fig. 1.  For comparison, we obtain 4.9 events for 
the sum of all backgrounds.  Hence, we expect to have sensitivity
in a significant area of the $\tilde{M}_{t}$-$\tilde{M}_{LSP}$
plane, and urge the Fermilab experiments to try this type of approach.

\begin{figure}
\hfil\psfig{figure=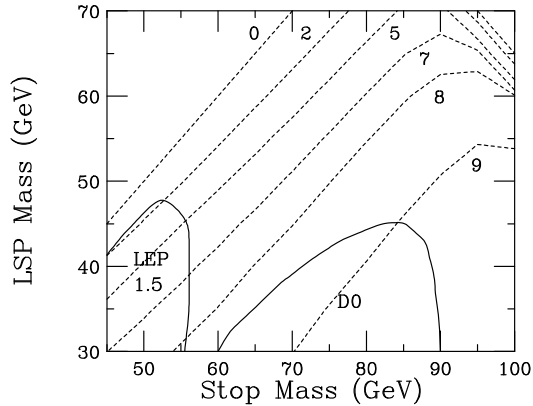,height=2.10in}
\caption{Predicted number of signal events in 100 pb$^{-1}$
passing all of the cuts in Table 1, and satisfying the condition
$\suwgt\ge6$, as a function of the stop and LSP masses,
assuming a $K$ factor of unity and 
${\cal B}(t \rightarrow \tilde{t}\tilde\chi_1^0)=50\%$.
The regions labelled D0 and
LEP 1.5 have been excluded by stop pair 
production searches.$^{4}$}
\end{figure}



\end{document}